\documentstyle[preprint,pra,aps,epsf,epsfig,floats,amssymb]{revtex}

\begin{document}
\draft


\title{Formation of electron-hole pairs in a one-dimensional random
  environment}

\author{Mark Leadbeater,$^1$ Rudolf A. R\"omer,$^2$ and Michael
  Schreiber$^2$}

\address{$^1$Max-Planck-Institut f\"ur Physik komplexer Systeme,
  D-01187 Dresden\\ $^2$Institut f\"{u}r Physik, Technische
  Universit\"{a}t Chemnitz, D-09107 Chemnitz}

\date{$Revision: 1.11 $, compiled \today}
\maketitle

\begin{abstract}
  We study the formation of electron-hole pairs for disordered systems
  in the limit of weak electron-hole interactions. We find that both
  attractive and repulsive interactions lead to electron-hole pair
  states with large localization length $\lambda_{2}$ even when we are
  in this non-excitonic limit.  Using a numerical decimation method to
  calculate the decay of the Green function along the diagonal of
  finite samples, we investigate the dependence of $\lambda_2(U)$ on
  disorder, interaction strength $U$ and system size.  Infinite sample
  size estimates $\xi_{2}(U)$ are obtained by finite-size scaling.
  The results show a great similarity to the problem of two
  interacting electrons in the same random one-dimensional potential.
\end{abstract}

\pacs{71.55.Jv, 72.15.Rn, 71.35.-y}

\narrowtext


%
%

It is well-known that randomness leads to localization of
non-interacting electrons and holes. This effect is especially strong
in one-dimensional (1D) systems \cite{LR85,AALR79} leading to complete
localization even for very small disorder \cite{KM93}. On the other
hand, in the limit of strong attractive Coulomb interaction, electrons
and holes will pair into excitons with large lifetimes.
In the present work, we concentrate on the intermediate problem of
weakly interacting electrons and holes (IEH) in a random environment.
The energy scales are such that the band width is larger than the
disorder $W$ which in turn is larger than the interaction strength $U$
and so we do not have bound excitons, but rather electron-hole pairs.
Such a problem is relevant for the proposed experimental verification
of the two-interacting particle (TIP) effect by optical experiments in
semiconductors \cite{G98}.
The TIP problem has recently attracted a lot of attention after
Shepelyansky \cite{S94,S96} argued that attractive as well as
repulsive onsite interactions between two bosons or fermions {\em in a
  single random potential} lead to the formation of particle pairs
whose localization length $\lambda_2 \propto U^2 \lambda_1^2$ is much
larger than the single-particle (SP) localization length $\lambda_1$.
His major prediction is that in the limit of weak disorder a pair of
particles will travel much further than a SP even for repulsive
interaction $U > 0$.

Here, we consider the effect of onsite interaction on a single
electron-hole pair, modelled by TIP in {\em two different 1D random
  potentials}. The Hamiltonian we consider is
\begin{eqnarray} 
H & = &
- t \sum_{n,m} \left( \vert n, m \rangle\langle n+1,m\vert + 
                      \vert n, m \rangle\langle n,m+1\vert + h.c. \right)
\nonumber \\
 & & \mbox{ }
+ \sum_{n, m} \vert n, m \rangle 
              \left( \epsilon^1_n + \epsilon^2_m + U\delta_{nm} \right) 
              \langle n,m\vert
\label{eq-ham}
\end{eqnarray} 
where for the case of TIP in different potentials, e.g. two electrons
on neighboring chains, or an electron and a hole on the same chain, we
have $\epsilon^1_n \neq \epsilon^2_n$. For simplicity, both
$\epsilon^1_n$ and $\epsilon^2_n$ are chosen randomly from the
interval $[-W/2,W/2]$.  In the following we call this situation the
IEH case.
We will show that the results for IEH are similar to the (standard)
TIP problem when both particles are in the same potential
$\epsilon^1_n =\epsilon^2_n$. In both cases $U$ is the on-site
interaction between the two particles.  We use hard-wall boundary
conditions and the hopping element $t\equiv 1$ sets the energy scale.
We note that this Hamiltonian corresponds to the 2D Anderson model if
we replace $\epsilon^1_n+\epsilon^2_m$ in (\ref{eq-ham}) by
$\tilde{\epsilon}_{nm}$ and choose $\tilde{\epsilon}_{nm}\in
[-W/2,W/2]$. In this case $U$ is an added on-site potential. This has
been recently used to test our numerical method by comparing with more
established methods valid for this model \cite{LRS98}.

To obtain our results we use a decimation method (DM)
\cite{LRS98,LW80}.  This involves replacing the full Hamiltonian by an
effective Hamiltonian for the doubly-occupied sites only. It should be
stressed that this method is exact and no approximations have to be
made in the decimation process. It is then possible, via a simple
inversion, to obtain the Green function matrix elements $\langle
1,1\vert {G_2}\vert M,M\rangle$ between doubly-occupied sites.  We
shall be focusing on the IEH localization length $\lambda_2$ obtained
from the decay of the transmission probability of IEH from one end of
the system to the other. This is defined \cite{OWM96} by
\begin{equation}
  {1\over\lambda_2} = - {1\over\vert M-1\vert} \ln\vert\langle
  1,1\vert { G_2}\vert M,M\rangle\vert.
\label{eq-lambda2}
\end{equation}
In order to reduce possible boundary effects, we compute $\lambda_2$
by considering the decay between sites slightly inside the sample.  We
present results for the band centre, i.e., energy $E=0$ for $14$
disorder values $W$ between $0.5$ and $7$, for 21 system sizes $M$
between $51$ and $251$, and 11 interactions strengths $U= 0, 0.1,
\ldots, 1.0$. For each triplet of parameters $(W,M,U)$ we average the
localization lengths $\lambda_2$ computed from the Green function
according to Eq.\ (\ref{eq-lambda2}) over 100 samples.

In Fig.\ \ref{fig-iehdm-l2_w}, we show the IEH results for $M= 201$.
Let us first turn our attention to the case $U=0$. As pointed out
previously \cite{SK97}, the TIP Green function ${ G_2}$ at $E=0$ is
given by a convolution of two SP Green functions ${ G_1}$ at energies
$E_1$ and $-E_1$. This implies that $\lambda_2 \approx \lambda_1/2$.
We have therefore included data for $\lambda_1/2$ in Fig.\ 
\ref{fig-iehdm-l2_w} which we have computed by a transfer matrix
method (TMM) \cite{CKM81} in 1D with $0.1\%$ accuracy.  Comparing
these results to the localization lengths $\lambda_2$ obtained from
the DM, we find that for $1 \leq W \leq 6$, the agreement between
$\lambda_2(U=0)$ and $\lambda_1/2$ is rather good and, contrary to TMM
results \cite{FMPW95,RS97,HZKM98}, there is no large artificial
enhancement at $U=0$. For smaller disorders $W < 1$, we have
$\lambda_2 \approx M/2$ so that it is not surprising that the Green
function becomes altered due to the finiteness of the chains
\cite{VRS97}.  This results in reduced values of $\lambda_2$.  It is
noticeable from these results, however, that the values of
$\lambda_2(U=0)$ are still slightly larger than $\lambda_1/2$.  This
is similar to TIP results \cite{SK97} and to a numerical convolution
of the SP Green functions calculated by exact diagonalisation
\cite{LRS98}.  For $U$ between $0.1$ and $1$ and $W \geq 1.2$ we have
found that the localization lengths $\lambda_2(U)$ are increased by
the onsite interaction as shown in Fig.\ \ref{fig-iehdm-l2_w}. For
$U\gg 1$, it is well-known that the interaction will split the single
TIP band into upper and lower Hubbard bands and, hence, we expect that
for large $U$ the enhancement of the localization length at $E=0$ will
vanish.  In Fig.\ \ref{fig-iehdm-l2_u} we present data for
$\lambda_2(U)/\lambda_2(0)$ for $U=-4, \ldots, 4$. We first observe
that at the band centre the enhancement is symmetric in $U$.  This is
why we usually only consider $U>0$ in agreement with the previous
arguments and calculations for the TIP case
\cite{LRS98,OWM96,SK97,FMPW95,RS97,HZKM98,VRS97,PS97,WWP98}. We have
checked that away from the band centre the enhancement is asymmetric
in $U$.  For small $|U|$, we see from Fig.\ \ref{fig-iehdm-l2_u} that
the localization length increases nearly linearly in $|U|$ with a
slope that is larger for smaller $W$ and we do not see any $U^2$
behavior as argued in Refs.\ \cite{S94,S96,I95}.  At large $|U|$ the
enhancement starts to decrease again.  For TIP, it has been suggested
\cite{WWP98} that there exists a duality for $U$ and $\sqrt{24}/U$ for
very large $|U|$.  The crossover between the two asymptotic regimes
should accor at $U_c = 24^{1/4}$. For IEH, we find that within the
accuracy of our data, we can argue for an agreement with the duality.
As for our TIP data \cite{LRS98}, we observe the best IEH agreement
with duality for $W=5$ but the maximum enhancement
$\max_{U}\left[\lambda_2(U)/\lambda_2(0)\right]$ still seems to depend
upon the disorder.

In order to overcome the problems with the finite chain lengths, we
construct finite-size scaling (FSS) curves for each $U$ and compute
from these scaling parameters which are the infinite-sample
localization lengths $\xi_2(U)$.  This method has been proven very
useful for the non-interacting case \cite{MK83} and recently for TIP
studies \cite{LRS98,SK97}.  In Fig.\ \ref{fig-iehdm-l2_m} we show the
raw data of the reduced IEH localization lengths $\lambda_2/M$ which
are to be scaled just as in the standard TMM \cite{MK83}. Note that
data for small $W$ are rather noisy and will thus most likely not give
very accurate scaling.  In order to set an absolute scale in the FSS
procedure, one usually fits the smallest localization lengths of the
largest systems to $\lambda_2/M = \xi_2/M + b (\xi_2/M)^2$ with $b$
small \cite{MK83}.  Due to numerical problems of estimating a small
localization length of the order of $1$ in a large system by Eq.\ 
(\ref{eq-lambda2}) we instead fit for each $U$ to the localization
length at $W=3$ and adjust the absolute scale of $\xi_2$ accordingly.
In Fig.\ \ref{fig-iehdm-fss} we show the resulting scaling curves
$\lambda_2/M = f(\xi_2 /M)$ for $U=0$, $0.2$ and $1$.  The above
mentioned numerical errors in the data at large $M$ and $W$ are
visible only in very small upward deviations from the expected $1/M$
behavior. The results are very similar to the TIP problem. It is
interesting to note that it is even possible to scale the present IEH
data together with the data previously obtained for the TIP case
\cite{LRS98}. From this more accurate scaling we compute the scaling
parameters $\xi_2$ which we show in Fig.\ \ref{fig-iehdm-xi2_w}.  The
power-law fits $\xi_2 \propto W^{-2\alpha}$ to the data with $W \in
[1,5]$ yield an exponent $\alpha$ which increases with increasing $U$
as shown in the inset of Fig.\ \ref{fig-iehdm-xi2_w}, e.g.,
$\alpha\approx 1.1$ for $U=0$ and $\alpha\approx 1.5$ for $U=1$.
Thus, although in Fig.\ \ref{fig-iehdm-l2_w} the $\lambda_2$ data at
$M=201$ nicely follows $\lambda_1/2$ for $U=0$, we nevertheless find
that after FSS with data from all system sizes, $\xi_2(0)$ as against
$\lambda_1/2$ still gives a slight enhancement.  Because of this we
will in the following compare $\xi_2(U>0)$ with $\xi_2(0)$ when trying
to identify an enhancement of the localization lengths due to
interaction. For comparison, the exponents obtained from the same fit
applied to the TIP problem \cite{LRS98} are also shown.  Note that, as
expected from Fig.\ \ref{fig-iehdm-l2_m}, FSS is not very accurate for
small $W$. Therefore, in what follows we shall only use $\xi_2$ values
obtained for $W\geq 1$. For data corresponding to $W<1$, we actually
used the extrapolated values of $\xi_2$ from the power-law fit to
continue the FSS curves of Fig.\ \ref{fig-iehdm-fss} to $W< 1$.

We now compare our IEH results with various fits proposed for TIP.
From an effective random matrix model \cite{S94,VRS97,PS97}
$\lambda_2\,\propto \, \lambda_1^{\beta}$ was obtained for large
values of $\lambda_1$.  To correct for smaller values of $\lambda_1$ a
more accurate expression was suggested \cite{PS97} to be $\lambda_2\,
\propto\, \lambda_1^{\beta}(1+c/\lambda_1)$. It is important that
$\beta$ in this work depends on $U$ and ranges from $1$ at small $U$
and very large $U$ to nearly $2$ for $U\approx t$.  As discussed above
we translate this fit function into $\xi_2(U) \propto \xi_2(0)^{\beta}
\left( 1 + \frac{c}{\xi_2(0)} \right)$. In Fig.\ \ref{fig-iehdm-ps} we
show respective data for disorders $W \in [1,6]$. The fits are good
and reflect in particular the deviations from a simple power-law
$\xi_2(U) \propto \xi_2(0)^\beta$ for small localization lengths. In
the inset we present the dependence of $\beta$ on $U$: we find $\beta
< 1.5$ for all $U$ values considered unlike Ref.\ \cite{PS97}. The
values obtained for the TIP case \cite{LRS98} are shown for
comparison.

In Ref.\ \cite{OWM96} the functional dependence of the TIP
localization lengths $\lambda_2 = \lambda_1/2 + c |U| \lambda_1^2$ has
been suggested. Taking instead of $\lambda_1/2$ the more suitable
$\xi_2(0)$ this can be translated as $ \xi_2(U) - \xi_2(0) \propto |U|
\xi_2(0)^2$.  In Fig.\ \ref{fig-iehdm-xi2_all} we plot
$[\xi_2(U)-\xi_2(0)]/|U|$ vs.\ $g(U)\xi_2(0)$ for $U\in[0.1,1]$ where
we have chosen $g(U)$ so that the data for different $U$ can be placed
on top of the $U=0.1$ data.  In the inset of Fig.\ 
\ref{fig-iehdm-xi2_all} we show that $g(U)$ starts to deviate from $1$
already for $U \geq 0.4$. Thus we see that the linear behavior in
$|U|$ found in Ref.\ \cite{OWM96} holds only for very small $U$ in
agreement with TIP.  We obtain a good fit to $[\xi_2(U)-\xi_2(0)]/|U|$
with a single exponent $\beta =1.6\pm 0.1$ instead of $2$.  This is
somewhat different from the fits for TIP which give $\beta\approx 2$
for small $\xi_2(0)<10$ and $\beta\approx 3/2$ for larger $\xi_2(0)$
\cite{LRS98}.  The reduction of the slope below $\beta=2$ may be due
to insufficient disorder averaging and thus an underestimation of the
FSS results.

In conclusion, we have presented detailed results for the localization
lengths of electron-hole pair states which may be realized
in the non-excitonic limit of optically excited semiconductor
heterostructures \cite{G98}.  We observe an increase of the
two-particle localization length due to onsite interaction in the band
centre. This suggests the formation of an electron-hole pair with
possibly enhanced transport properties. We emphasize that our results
apply to the non-excitonic limit with bandwidth larger than $W >U$.
We have fitted our data to various suggested models with varying
success.  The results are all similar to the standard TIP problem in a
single random potential and thus we conclude that the
case of an interacting electron-hole pair is very close to the
TIP problem \cite{LRS98}.

We thank E.\ McCann, J.\ E.\ Golub, O.\ Halfpap and D.\ Weinmann for
useful discussions.  R.A.R.\ gratefully acknowledges support by the
Deutsche Forschungsgemeinschaft through SFB 393.

%
%

%
%

\newcommand{\figwidth}{0.8\columnwidth}
\newcommand{\figheight}{0.8\textheight}

\pagebreak

\begin{figure}[th]
\centerline{\epsfig{figure=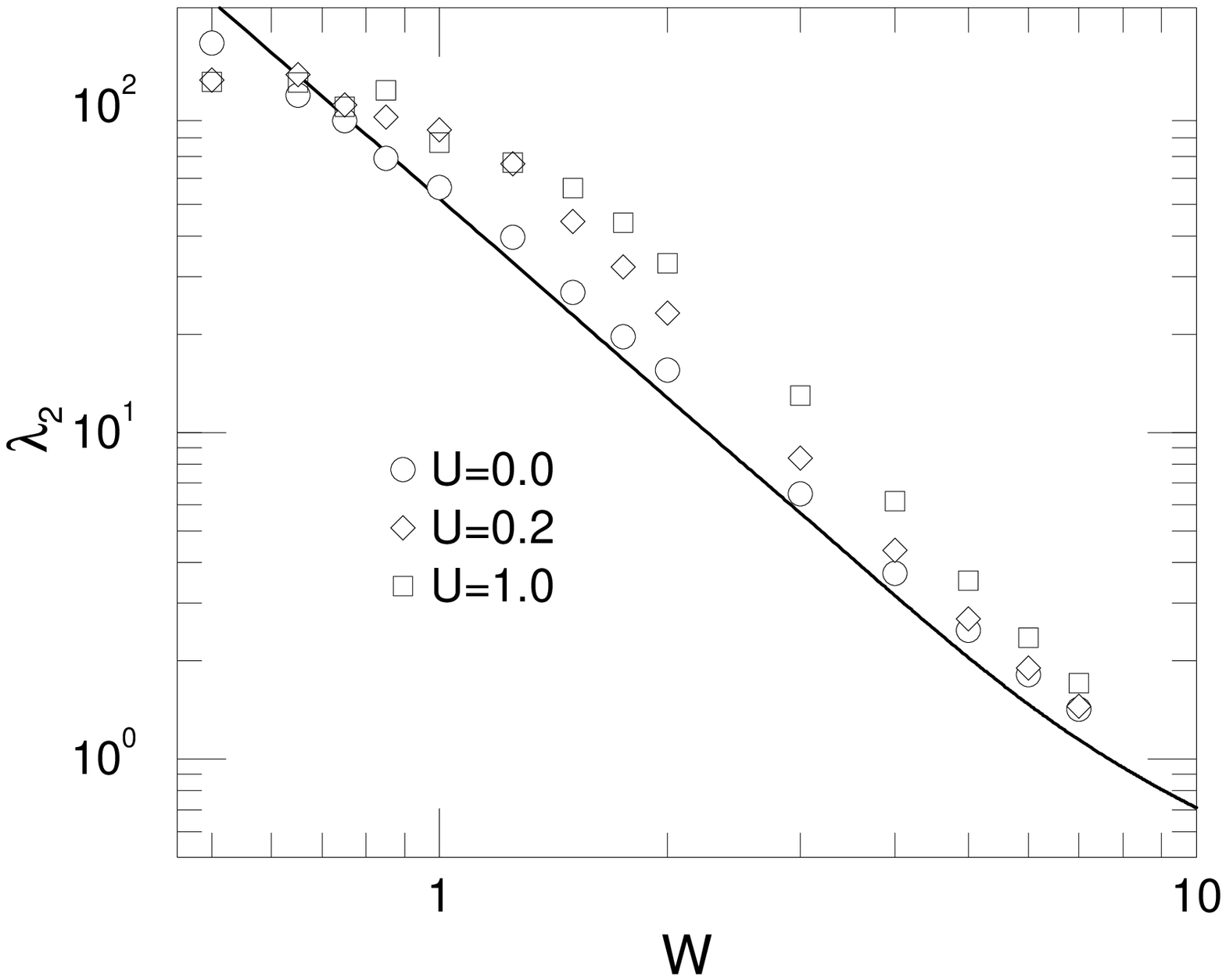,width=\figwidth} }
\caption{\label{fig-iehdm-l2_w}
  Localization length $\lambda_2$ at energy $E=0$ for system size
  $M=201$ and different interaction strengths. The thick solid line
  represents data for $\lambda_1/2$.  }
\end{figure}

\begin{figure}[th]
\centerline{\epsfig{figure=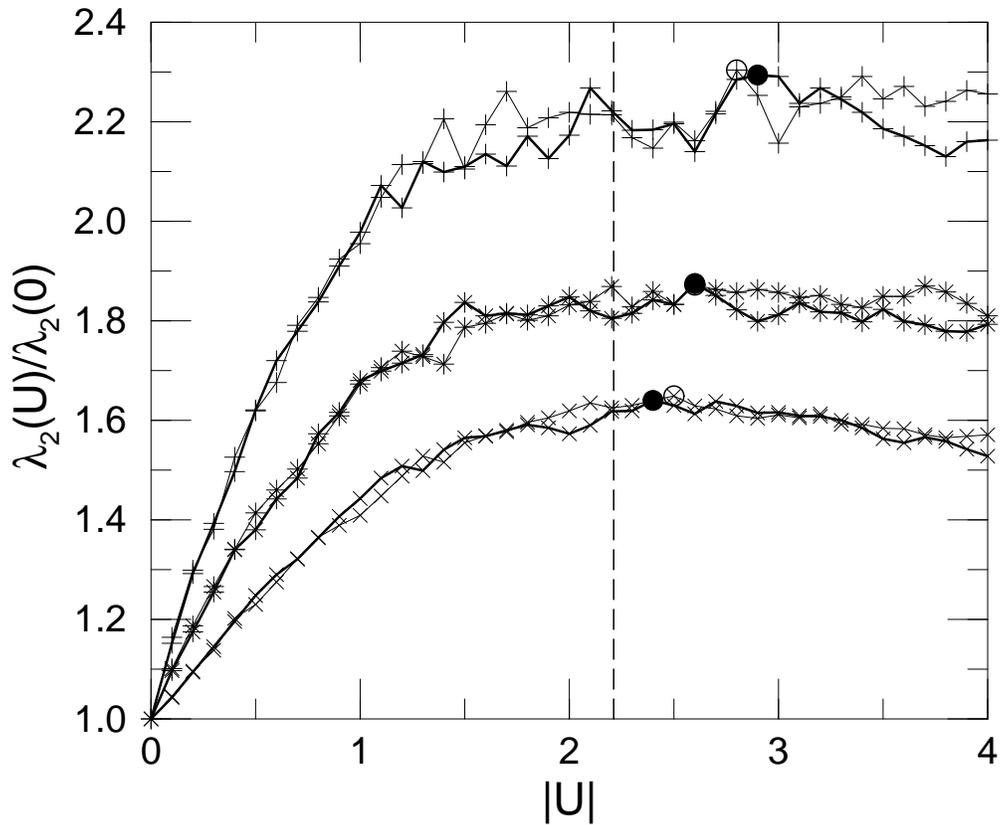,width=\figwidth} }
\caption{\label{fig-iehdm-l2_u}
  Enhancement $\lambda_2(U)/\lambda_2(0)$ as a function of interaction
  strength $U$ at $E=0$ for disorder $W=3$ ($+$), $4$ ($*$), and
  $5$ ($\times$) and $M=201$.  The thick (thin) lines indicate data
  for $U>0$ ($U<0$), full (open) circles denote the maximum for each
  disorder. The dashed line marks $U_c=24^{1/4}$.}
\end{figure}

\begin{figure}[th]
\centerline{\epsfig{figure=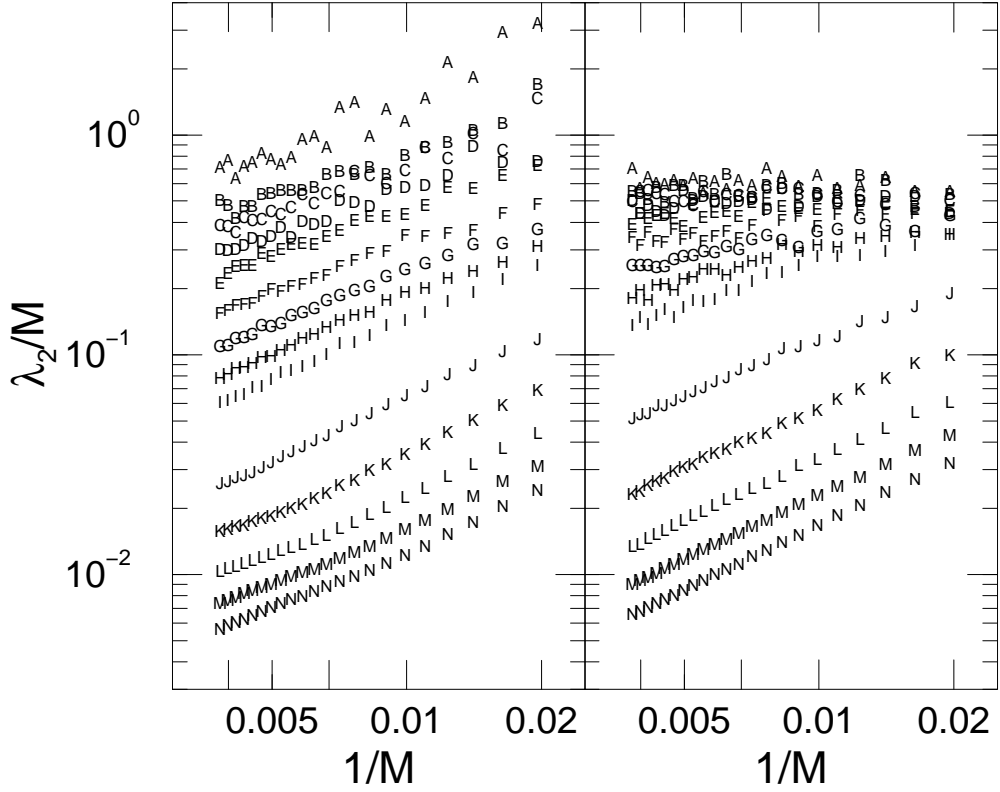,width=\figwidth} }
\caption{\label{fig-iehdm-l2_m}
  Reduced localization lengths $\lambda_2/M$ for $U=0$ (left) and
  $U=1$ (right) for $14$ disorders between $0.5$ (A) and $7$ (N) as in
  Fig.\ \protect\ref{fig-iehdm-l2_w}.  }
\end{figure}

\begin{figure}[th]
\centerline{\epsfig{figure=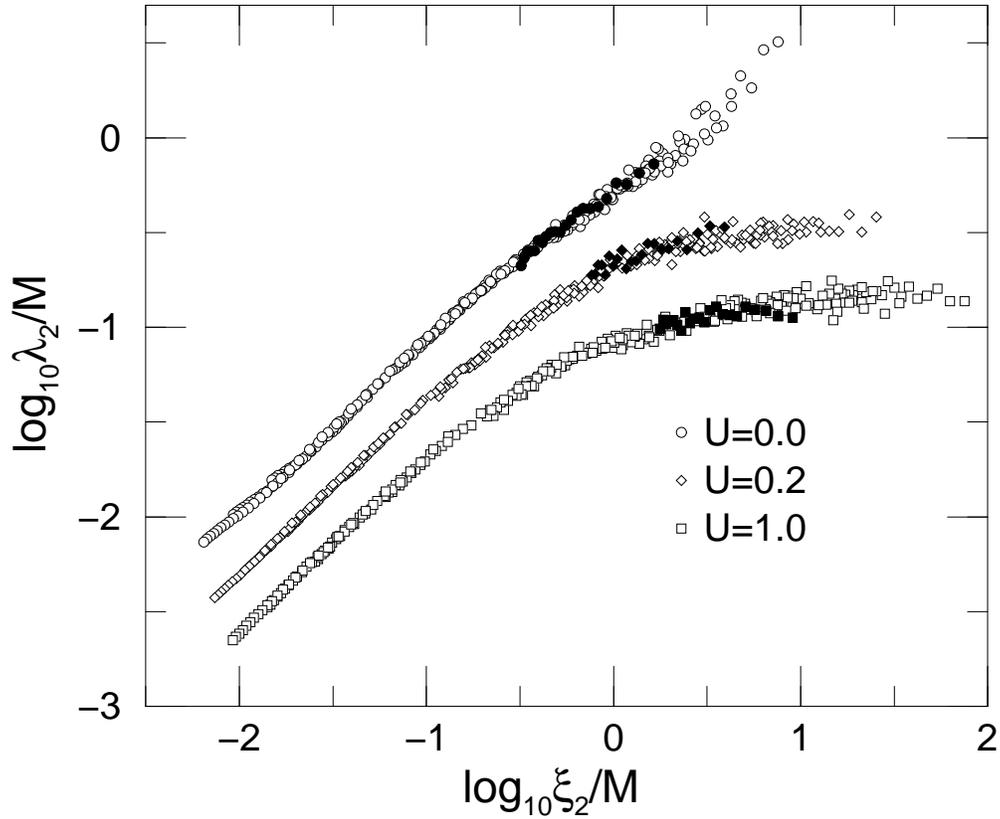,width=\figwidth} }
\caption{\label{fig-iehdm-fss}
  Finite-size scaling plot of the reduced localization lengths
  $\lambda_2/M$ for various $U$. The data for $U=0.2$ ($1$) have been
  divided by 2 (4) for clarity. Data corresponding to $W=1$ are
  indicated by filled symbols. }
\end{figure}

\begin{figure}[th]
\centerline{\epsfig{figure=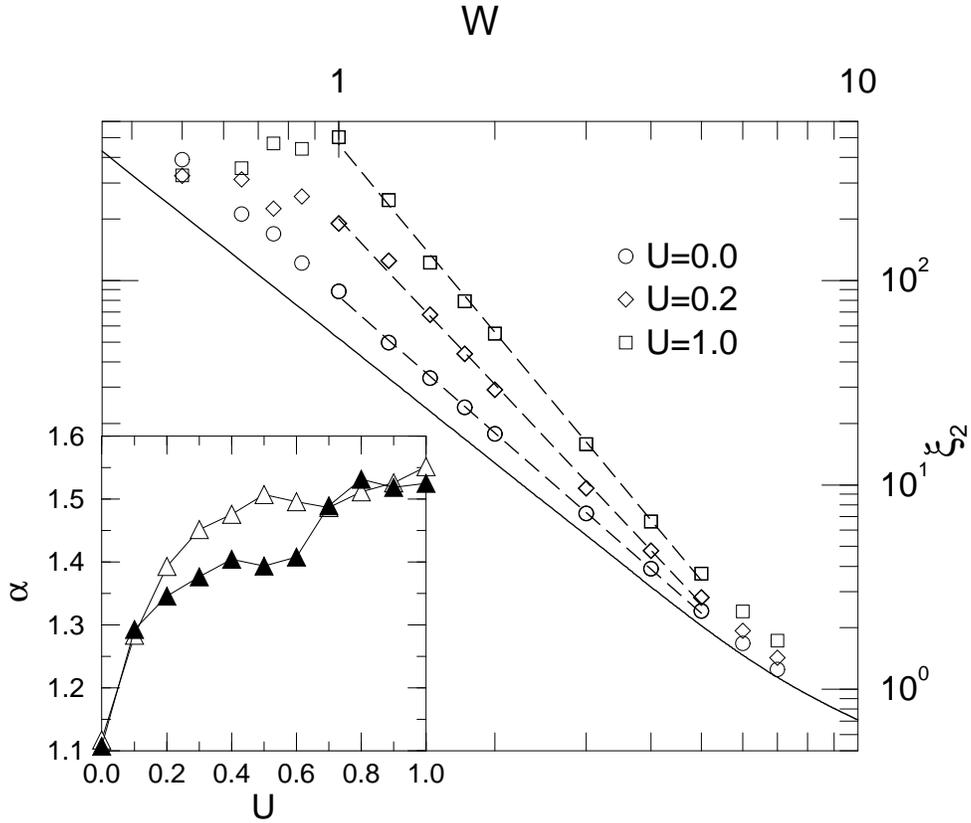,width=\figwidth} }
\caption{\label{fig-iehdm-xi2_w}
  Localization lengths $\xi_2$ after FSS for various $U$. The solid
  line represents $\lambda_1/2$, the dashed lines indicate power-law
  fits. Inset: Exponent $\alpha$ obtained by fitting $\xi_2 \propto
  W^{-2\alpha}$ to the data for each $U$ (filled symbols).  The result
  for the standard TIP problem is also shown (open triangles).}
\end{figure}

\begin{figure}[th]
\centerline{\epsfig{figure=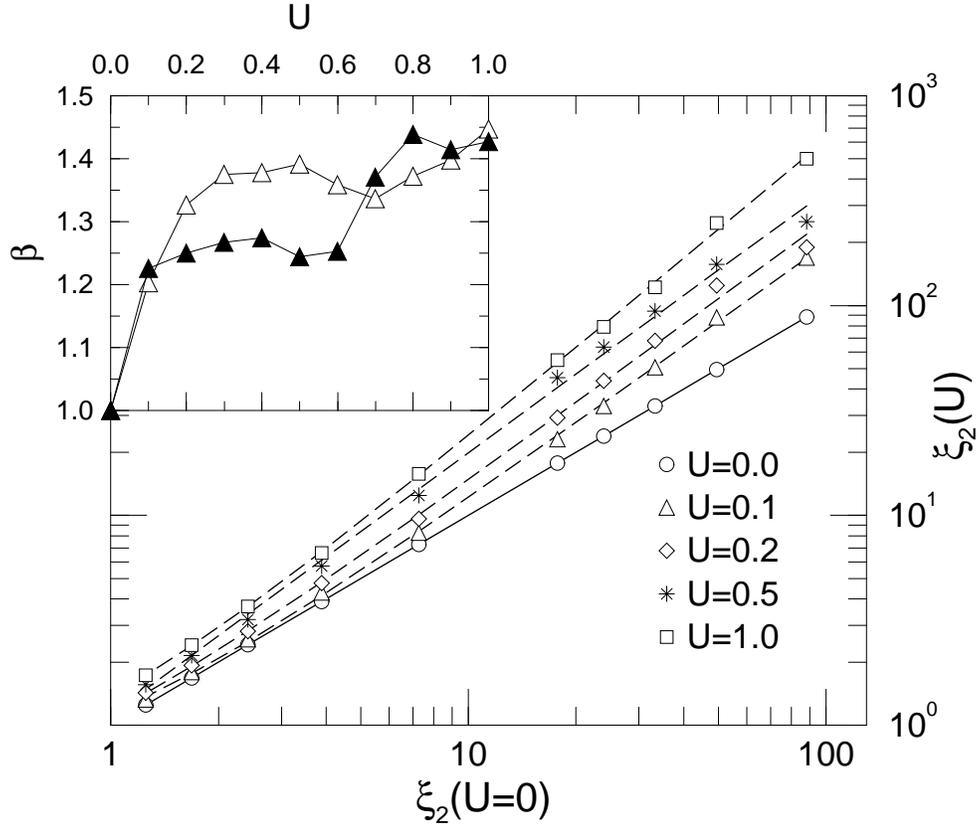,width=\figwidth} }
\caption{\label{fig-iehdm-ps} 
  $\xi_2(U)$ after FSS for various $U$ plotted versus $\xi_2(0)$. The
  lines are fits $\xi_2(U)\propto\xi_2(0)^{\beta}[1+c/\xi_2(0)]$.
  Inset: Exponent $\beta$ from the fits (filled symbols) and for the
  standard TIP problem (open triangles).}
\end{figure}

\begin{figure}[th]
\centerline{\epsfig{figure=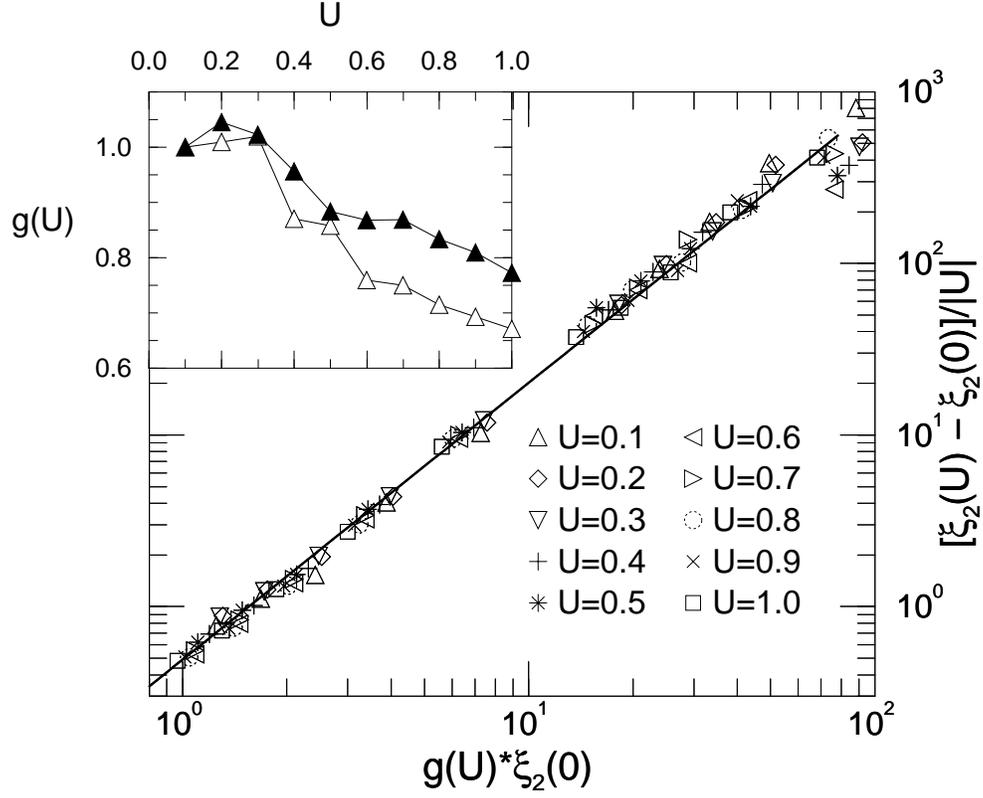,width=\figwidth} }
\caption{\label{fig-iehdm-xi2_all}
  Enhancement of the localization lengths $[\xi_2(U)-\xi_2(0)]/|U|$
  plotted for $W\in[1,7]$ versus $g(U)\xi_2(0)$ where $g(U)$ was
  obtained by a mean-least-squares fit procedure to make all the data
  compatible with $g(0.1)=1$. The straight line is the curve
  $\xi_2(U)= \xi_2(0)+0.49|U|[g(U)\xi_2(0)]^{1.61}$.  Inset: Behaviour
  of $g(U)$ for IEH (filled symbols) and for TIP \protect\cite{LRS98}
  (open triangles).}
\end{figure}

\end{document}